# Sub-nanoscale Temperature, Magnetic Field and Pressure sensing with Spin Centers in 2D hexagonal Boron Nitride

Andreas Gottscholl[1], Matthias Diez[1], Victor Soltamov[1,2], Christian Kasper[1], Andreas Sperlich[1], Mehran Kianinia[3,4], Carlo Bradac[5], Igor Aharonovich[3,4], Vladimir Dyakonov[1,*]

[1] Experimental Physics 6 and Würzburg-Dresden Cluster of Excellence ct.qmat, Julius Maximilian University of Würzburg, 97074 Würzburg, Germany
[2] Ioffe Institute, St. Petersburg 194021, Russia
[3] School of Mathematics and Physical Sciences, University of Technology Sydney, Ultimo, NSW 2007, Australia
[4] Centre of Excellence for Transformative Meta-Optical Systems, University of Technology Sydney, Ultimo, NSW 2007, Australia
[5] Department of Physics & Astronomy, Trent University, 1600 West Bank Dr., Peterborough 28 ON, K9J 0G2, Canada

**Spin defects in solid-state materials are strong candidate systems for quantum information technology and sensing applications. Here we explore in details the recently discovered negatively charged boron vacancies ($V_B^-$) in hexagonal boron nitride (hBN) and demonstrate their use as atomic scale sensors for temperature, magnetic fields and externally applied pressure. These applications are possible due to the high-spin triplet ground state and bright spin-dependent photoluminescence (PL) of the $V_B^-$. Specifically, we find that the frequency shift in optically detected magnetic resonance (ODMR) measurements is not only sensitive to static magnetic fields, but also to temperature and pressure changes which we relate to crystal lattice parameters. Our work is important for the future use of spin-rich hBN layers as intrinsic sensors in heterostructures of functionalized 2D materials.**

Spin defects in three-dimensional (3D) wide band-gap semiconductors have extensively been utilized in both fundamental and practical realizations in quantum science. The most prominent systems are the nitrogen-vacancy (NV) center in diamond[1] and various types of spin defects in silicon carbide (SiC) (divacancy and silicon vacancy)[2,3]. These systems reveal optically detected magnetic resonance (ODMR), which allows for polarization, manipulation and optical readout of their spin state and consequent mapping of external stimuli (magnetic/electric field, temperature, pressure, etc.) onto it[4-6]. A variety of reports have demonstrated outstanding nanoscale sensing applications of NV-centers (particularly NV⁻) in physics and biology including detection of individual surface spins[7] and nanothermometry in living cells[8,9]. However, NV-centers in diamond and spin centers in SiC possess intrinsic limitations. The three-dimensional nature of the material makes it challenging to position the spin-defects close as to the sample surface, and thus, to the object/quantity to be sensed. Furthermore, the proximity to the surface deteriorates their spin coherence properties and hinder their sensitivity as nano-sensors[10].

A remedy to these limitations may be provided by recently discovered defects in layered materials. One of the most prominent stackable 2D materials is hexagonal boron nitride (hBN) which hosts a large variety of atom-like defects including single photon emitters[11-14]. Spin carrying defects have been theoretically predicted and experimentally confirmed in hBN[15-19]. Currently, the most understood defect is the negatively-charged boron vacancy center ($V_B^-$)[20], which can be readily created by neutron irradiation, ion implantation or femtosecond laser pulses[21,22]. Due to its spin-optical and properties, the $V_B^-$ center is proving to be a promising candidate system for quantum information and nanoscale quantum sensing applications and has thus expanded the already large suite of unique features displayed by 2D materials[13].

The recently identified $V_B^-$ in hBN displays a photoluminescence (PL) emission band around 850 nm and has been found to be an electronic spin-triplet (S=1) system with a ground state zero-field splitting (ZFS) $D_{gs}/h \cong$ 3.5 GHz between its spin sublevels $m_s = 0$ and $m_s = \pm 1$[16]. Here, we study the effect of external stimuli on the defect's properties and demonstrate its suitability for sensing temperature, pressure (as lattice compression) and magnetic fields. Notably, our experiments show that the resolution and range of operation of the hBN $V_B^-$ center is competitive or exceeding those of similar defect-based sensors[23].

The results presented in this work were obtained on single-crystal hBN. The $V_B^-$ centers were generated in the sample via neutron irradiation (≈2.3×10¹⁸ n·cm⁻²), as described elsewhere[16,20]. The hBN single crystalline sample consists of a stack of a few thousand mono layers. The distance between two adjacent layers is $c \cong$ 6.6 Å, while the in-plane distance between two identical atoms is a $\cong$ 2.5 Å (Fig. 1a). As shown by temperature dependent X-ray data[24], the hBN lattice undergoes highly anisotropic thermal expansion with $c$ and a changing in opposite directions, i.e., while $c$ decreases with cooling, a increases, as schematically shown in Fig. 1b. This crystallographic feature can be used to monitor local temperature variations optically, via ODMR, since the temperature-driven compression/expansion of the lattice parameters a and $c$ causes a direct change in the ZFS parameter $D_{gs}$ of the triplet ground state Spin-Hamiltonian[25]. Figure 1c shows continuous wave (cw)

ODMR measurements for three different temperatures, with (dark blue) or without (cyan) an external magnetic field $B$ applied. At room temperature and in the absence of the external magnetic field, the ODMR spectrum of the $V_B^-$ shows two resonances ($\nu_1, \nu_2$) centered symmetrically around $\nu_0$, which corresponds to the ZFS parameter $D_{gs}/h = \nu_0 = 3.48$ GHz with the splitting due to the non-zero off-axial ZFS parameter $E_{gs}/h \cong 50$ MHz. When applying an external static magnetic field $B$, $\nu_1$ and $\nu_2$ split further following:

$$\nu_{1,2} = D_{gs}/h \pm (1/h)\sqrt{E_{gs}^2 + (g\mu_B B)^2}. \tag{1}$$

Here, h is Planck's constant, g is the Landé factor and $\mu_B$ is the Bohr magneton. The separation of the two resonances $\nu_1$ and $\nu_2$ can clearly be seen in Fig. 1c (dark blue traces). The visible substructure in both ODMR peaks is due to hyperfine coupling of the electron S=1 spin system (negatively charged boron vacancy) with three equivalent nearest nitrogen atoms, each possessing nuclear spin $I = 1$ for the most abundant $^{14}$N isotope (99.63%). In total, seven hyperfine peaks can be resolved, whose relative separations are temperature and magnetic field independent. A closer look at Fig. 1c reveals that cooling down the sample results in a shift of the ODMR peaks to higher frequencies. Thus, the dependencies of the ODMR spectrum on temperature and magnetic fields can provide a basis for the use of the $V_B^-$ center as a thermometer and magnetometer at the sub-nanoscale.

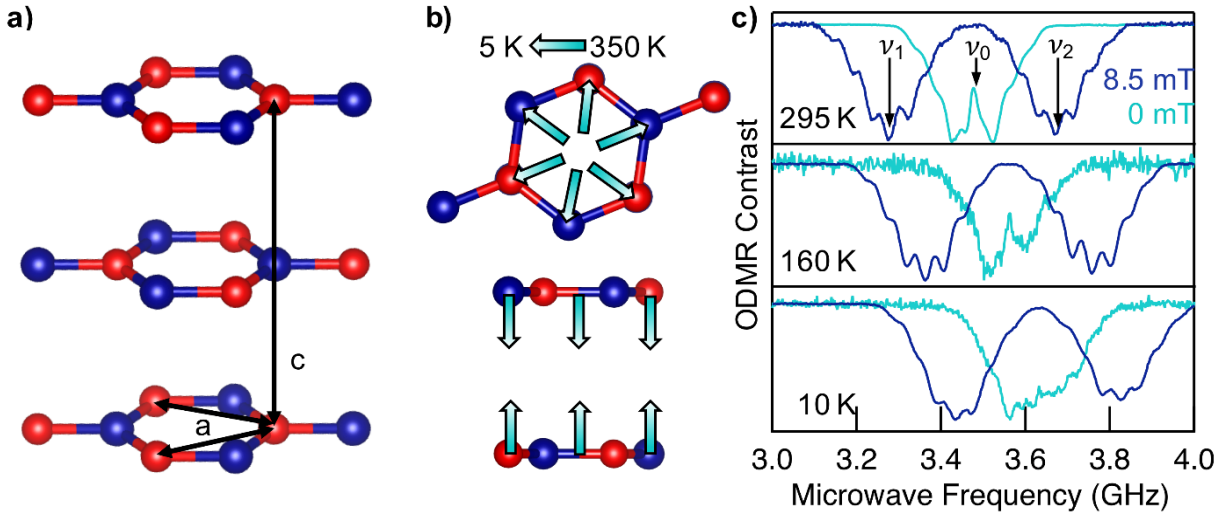

*Figure 1. a) Schematic of the hexagonal boron nitride (hBN) crystal, its hexagonal structure with alternating boron (red) and nitrogen (blue) atoms and the lattice constants* a *and* c*. b) Lattice contraction and expansion due to temperature variation, according to crystallographic data[24]. c) cw ODMR spectra measured with (dark blue) and without (cyan) external magnetic field at different temperatures T = 295 K, 160 K and 10 K. Lowering of the temperature causes the resonances $\nu_0, \nu_1$ and $\nu_2$ to shift to larger microwave frequencies indicating an increase of the zero-field splitting $D_{gs}$.*

**Temperature Sensing**

The observed shift of the resonances to higher frequency values (Fig. 1c) is independent of the applied magnetic field and is solely due to a reduction of the ZFS parameter $D_{gs}$. Over the temperature range 295 – 10 K, $D_{gs}$ undergoes a variation $\Delta D_{gs} \cong 195$ MHz. This is a relatively large change compared to analogous spin systems in 3D materials (≈30-fold). For instance, the NV⁻ center in diamond exhibits a shift $\Delta D_{gs} \cong 7$ MHz [25], while the $D_{gs}$ of $V_{Si}$ in SiC is almost constant over the same range. Only more complex spin defects such as Frenkel defects ($V_{Si}$-$Si_i$) in SiC display a comparably strong effect ($\Delta D_{gs} \cong 300$ MHz) [5].

To quantify this temperature-induced shift of the ground state triplet energy-levels we combine temperature- and magnetic field-dependent ODMR measurements. Figure 2 summarizes the shift of $D_{gs}$ in the ODMR spectrum as a function of temperature both, in the presence (a) and absence (b) of an external magnetic field. In Fig. 2a, an external magnetic field of 8.5 mT is applied. A monotonic, nearly linear increase of the resonance frequencies associated to a change in the zero-field splitting parameter $D_{gs}$ can be observed for temperatures down to 50 K. Zero-field ODMR (Fig. 2b) shows the same behavior. From Figs. 2a, b we now extract the ZFS values $D_{gs}/h$ and plot them against temperature (Fig. 2c). Both temperature dependencies, represented by dark blue and cyan diamonds, match perfectly and thus confirm that the temperature scaling of $D_{gs}$ is indeed independent of the magnetic field.

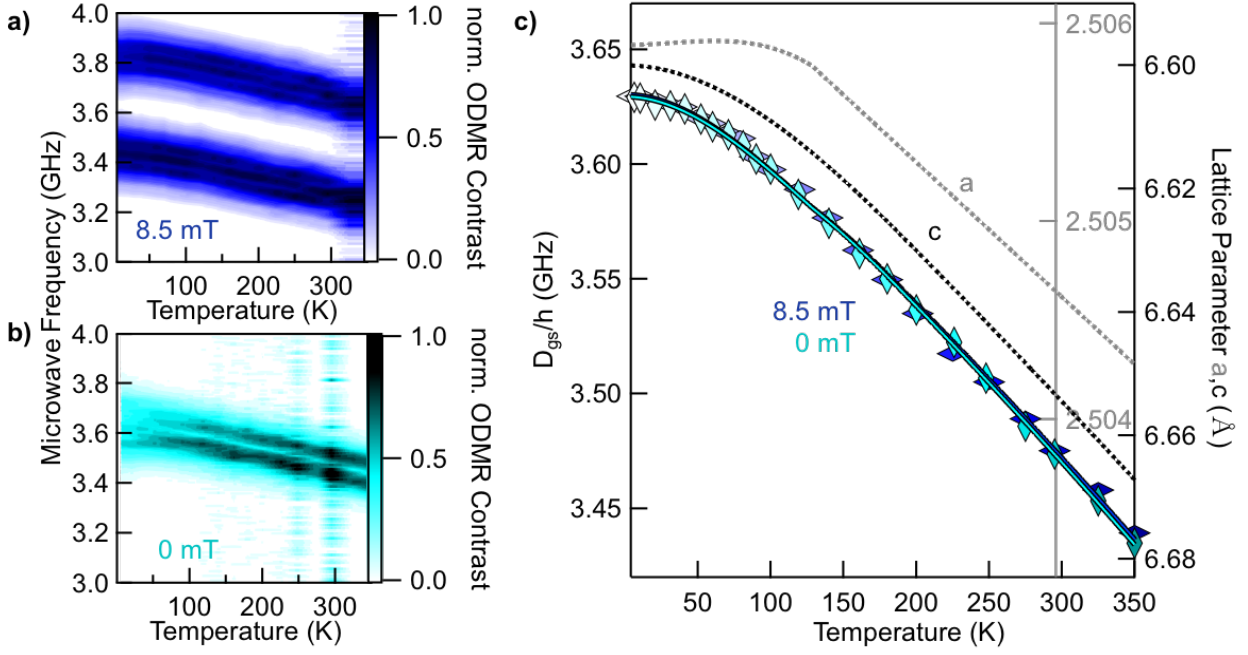

***Figure 2.*** *Temperature dependence of the ODMR spectrum of $V_B^-$. **a), b)** Color maps represent the peak positions of the normalized ODMR spectrum for different temperatures in an external magnetic field of (a) $B = 8.5$ mT and (b) $B = 0$ mT. **c)** ZFS parameter $D_{gs}/h$ obtained from a) (blue diamonds) and b) (cyan diamonds) vs. temperature. The monotonic increase while lowering the sample temperature is unaffected by the magnetic field. The data can be fitted using Eq. (2) describing the temperature-dependent change of the lattice parameters a (grey dotted line[24]) and c (black dotted line[24]) that are also plotted in (c). The fits are shown as solid lines (blue for 8.5 mT and cyan for 0 mT) on top of the diamonds and reproduce the temperature dependence perfectly.*

The temperature dependency can be explained by considering the change in the delocalization of the spin-defect wave function delocalization due to temperature-induced structural deformations of the crystal lattice. This is consistent with e.g. the case of the NV- center in diamond[25,26]. The latter shows a linear behavior for the shift of the ZFS associated to the relative change of the lattice constant $\eta$ is observed[26]: $\Delta D_{gs}(\eta)/h = \theta\, \eta(T)$, where $\theta$ is the proportionality factor, explicitly written as $dD/d\eta$ and $\eta(T)$ is the relative change of the lattice parameter. Applying the same concept to hBN with its two lattice parameters a and c results in the equation:

$$D_{gs}(\eta_a, \eta_c) = D_{gs,295K} + \Delta D_{gs,a} + \Delta D_{gs,c} = D_{gs,295K} + \theta_a \eta_a h + \theta_c \eta_c h \qquad (2)$$

Here, $D_{gs}/h$ is the experimentally measured ZFS frequency. $D_{gs,295K}/h$ = 3.48 GHz is the ZFS at T = 295 K that we choose as reference. $\Delta D_{gs,a}$ and $\Delta D_{gs,c}$ are the frequency shifts induced by the relative changes in a and c, $\eta_a(T)$ and $\eta_c(T)$ are relative changes of a and c, respectively (see also Eqs. (6,7) below). The temperature-dependent lattice parameters a(T) and c(T) for hBN were determined in Ref. 24 and are plotted in Fig. 2c in addition to the ODMR data. The proportionality factors $\theta_a$ and $\theta_c$ are the significant parameters that connect lattice deformation and ZFS and will be derived from the experimental data in the following. To do so, Eq. (2) is fitted to the experimentally measured ZFS $D_{gs}$, as shown in Fig. 2c. The fit perfectly reproduces the experimental data, which highlights the remarkable linear response of the resonance frequency to changes of the lattice constants in this case due to temperature.

Figure 3 shows the relationship between temperature-dependent lattice parameters and ZFS for a magnetic field of $B = 8.5$ mT. By inserting the crystallographic data for the hBN lattice parameters[24] into Eq. (2), we obtain a surface with respective slopes $\theta_a$ and $\theta_c$, as shown in Fig. 3d (grey).

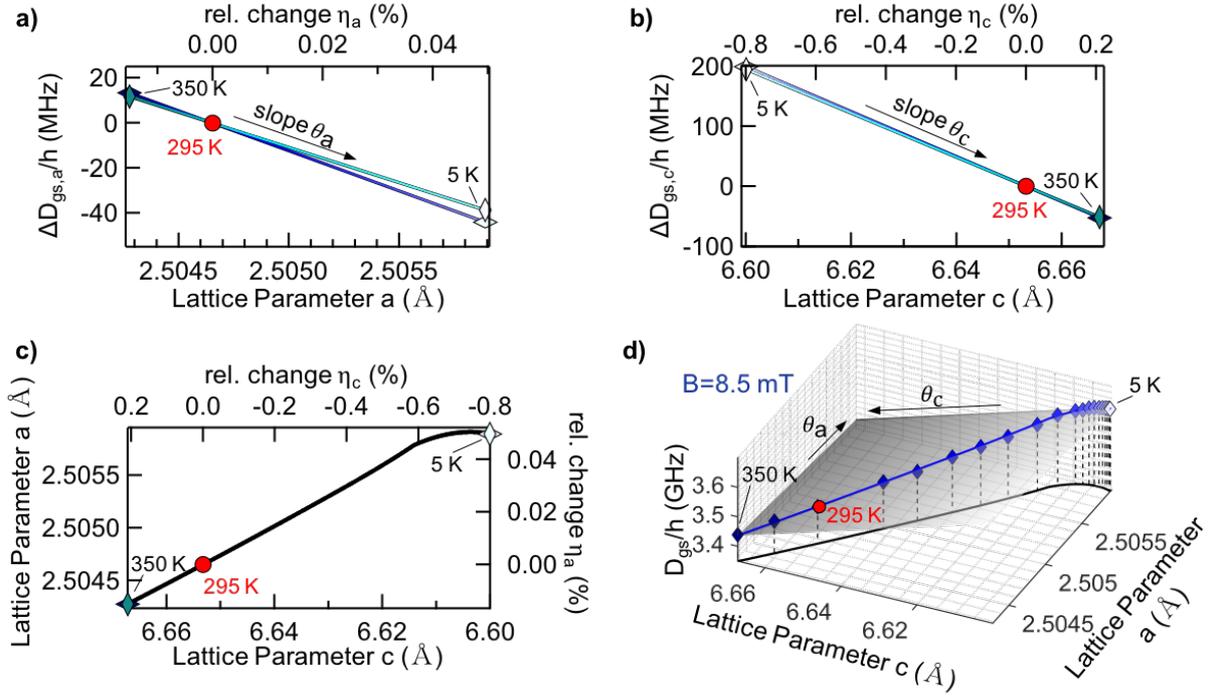

*Figure 3.* Zero-field splitting dependence on the lattice parameters a and c. Eq. (2) is fitted (solid lines) to the experimental data displayed as diamonds. **a)** Change of ZFS $\Delta D_{gs,a}$ caused by the temperature-dependent lattice parameter a. **b)** Change of ZFS $\Delta D_{gs,c}$ caused by temperature-dependent lattice parameter $c$. **c)** Comparison of lattice parameters in the temperature range 5 K – 350 K. **d)** Combined 3-dimensional representation of plots *a-c)*. Fitting Eq. (2) to the 3-dimensional data set $D_{gs}(a,c,T)$ allows estimating the slopes $\theta_a$ and $\theta_c$ (shown also Figs 3a, b). The ZFS reference temperature T = 295 K is marked by the red dot. The assignment of colors of fitting lines (blue for 8.5 mT and cyan for 0 mT) is the same as in Fig. 2.

We extract the values for $\theta_a$ of $(-84 \pm 15)$ GHz and $(-78.2 \pm 8.8)$ GHz or $\theta_c$ of $(-24.32 \pm 0.59)$ GHz and $(-24.6 \pm 1.0)$ GHz at B = 0 mT and 8.5 mT, respectively. The out-of-plane $\theta_c$ can be determined more precisely, since the relative change of lattice parameter $c$ is one order of magnitude larger. The values coincide within the experimental error and can be combined as:

$$\theta_a = (-81 \pm 12) \text{ GHz} \quad (3)$$
$$\theta_c = (-24.5 \pm 0.8) \text{ GHz} \quad (4)$$

Remarkably, the ratio $\theta_a/\theta_c \approx 3.3$, which means the influence of the lattice distortion on the ZFS in-plane is at least three times stronger than the influence of the interplanar distance on the same, indicating a localization of the $V_B^-$ spin density in the plane as predicted by the theory[17].

Finally, we propose a polynomial which allows a direct determination of $D_{gs}$ from the sample temperature $T$:

$$D_{gs}(T) = h \sum_k A_k T^k \quad (k = 0,1,2,3) \quad (5)$$

where T is the temperature, $h$ is Planck's constant, $k$ is an integer, and the polynomial coefficients $A_k$ are summarized in Table 1 for different temperature ranges. To obtain the coefficients $A_k$, the essential step is to determine the relative changes of the lattice parameters from the crystallographic data[24] by using the Equations (6) and (7):

$$\eta_a(T) = \frac{a(T) - a(295\,K)}{a(295\,K)} \quad (6)$$

$$\eta_c(T) = \frac{c(T) - c(295\,K)}{c(295\,K)} \quad (7)$$

| Temperature range (K) | $A_0$ (GHz) | $A_1$ (MHz K$^{-1}$) | $A_2$ (kHz K$^{-2}$) | $A_3$ (Hz K$^{-3}$) |
|---|---|---|---|---|
| 5-128 | 3.6367 | 0 | -4.4308 | 11.468 |
| 128-189 | 3.6109 | 0.22839 | -3.8805 | 5.522 |
| 189-350 | 3.6664 | -0.55659 | -0.2383 | 0 |

**Table 1.** *Calculated polynomial coefficients $A_k$ for Eq. 5. The three different temperature regions arise from the sectionally defined polynomial for the temperature- dependent lattice parameters a and c [24]. The exact procedure of calculation can be found in the Supporting Information.*

**Pressure Sensing**

The observation that a temperature-induced change in the lattice parameters directly results in a shift of the ZFS $D_{gs}$ leads to the consideration of utilizing the $V_B^-$ center also as a sensor, for externally applied in-plane or out-of-plane pressure. For a first-order estimation we assume an isothermal system without shear strain and derive the perspective sensitivity based on reported elastic moduli for hBN crystals[27,28]. In cartesian coordinates, the pressure vector $\sigma_{xyz}$ is given by the elastic moduli tensor $C$ multiplied with the relative change of the lattice parameters $\eta_{xyz}$:

$$\begin{pmatrix}\sigma_x \\ \sigma_y \\ \sigma_z\end{pmatrix} = \begin{pmatrix}C_{11} & C_{12} & C_{13} \\ C_{12} & C_{11} & C_{13} \\ C_{13} & C_{13} & C_{33}\end{pmatrix}\begin{pmatrix}\eta_x \\ \eta_y \\ \eta_z\end{pmatrix} \qquad (8)$$

The reported elastic moduli for hBN are: $C_{11}$ = (811±12) GPa, $C_{12}$ = (169±24) GPa, $C_{13}$ = (0±3) GPa and $C_{33}$ = (27±5) GPa[27]. The hBN lattice parameters a and c are oriented along the y- and z-direction, respectively. This simplifies Eq. (8) by incorporating $\eta_{a,c}$ and removing $C_{13}$, which is 0:

$$\begin{pmatrix}\sigma_x \\ \sigma_y \\ \sigma_z\end{pmatrix} = \begin{pmatrix}\eta_a\left(\tfrac{2}{3}C_{11}+\tfrac{1}{3}C_{12}\right) \\ \eta_a\left(\tfrac{2}{3}C_{12}+\tfrac{1}{3}C_{11}\right) \\ \eta_c C_{33}\end{pmatrix} \qquad (9)$$

This can be rewritten to obtain $\eta_{a,c}$ directly:

$$\eta_a = \sigma_x/\left(\tfrac{2}{3}C_{11}+\tfrac{1}{3}C_{12}\right) = \sigma_y/\left(\tfrac{2}{3}C_{12}+\tfrac{1}{3}C_{11}\right) \qquad (10)$$
$$\eta_c = \sigma_z/C_{33} \qquad (11)$$

Substituting these relationships into Eq. (2) yields the ZFS as a function of the applied pressure:

$$D_{gs}(\sigma_x,\sigma_y,\sigma_z) = D_{gs,295K} + \Delta D_{gs,x} + \Delta D_{gs,y} + \Delta D_{gs,z} = D_{gs,295K} + \frac{\theta_a\sigma_x h}{\left(\tfrac{2}{3}C_{11}+\tfrac{1}{3}C_{12}\right)} + \frac{\theta_a\sigma_y h}{\left(\tfrac{2}{3}C_{12}+\tfrac{1}{3}C_{11}\right)} + \frac{\theta_c\sigma_z h}{C_{33}} \qquad (12)$$

Based on our estimates for $\theta_{a,c}$ in Eqs. (3,4) and the reported elastic moduli[27], we obtain the sensitivity to measure the ZFS shifts for each direction of the applied pressure:

$$\Delta D_{gs,x} = \sigma_x h(-0.136 \pm 0.028)\tfrac{Hz}{Pa} \qquad (13)$$
$$\Delta D_{gs,y} = \sigma_y h(-0.212 \pm 0.052)\tfrac{Hz}{Pa} \qquad (14)$$
$$\Delta D_{gs,z} = \sigma_z h(-0.91 \pm 0.20)\tfrac{Hz}{Pa} \qquad (15)$$

Consequently, we find that the ZFS shift $\Delta D_{gs,xyz}$ is directly associated with external compression of the hBN lattice and therefore $V_B^-$ can be utilized as a pressure sensor. Remarkably, the out-of-plane sensitivity along the $c$-axis is much higher due to the small $C_{33}$ coefficient. This makes this type of sensor particularly useful to measure vertical indentation in 2D heterostructures.

**Magnetic Field Sensing**

As shown in Fig. 2a, the two resonant transitions $\nu_{1,2}$ are equally separated with respect to $\nu_0$ over the entire temperature range between 5 K and 350 K. It should be pointed out that the magnetic field sensing is based on the g-factor, which is independent of the lattice parameters. In Figure 4, we demonstrate the principle suitability of a $V_B^-$ center in hBN for magnetic field sensing, where we show the resonant microwave transitions $v_1$ and $v_2$ over a broad range ($0 - 3500 \text{ mT}$) and exemplarily simulated for two distant temperatures, T = 295 K (dark blue) and T = 5 K (light blue). For a magnetic field applied in the $c$-direction of the hBN lattice, the behavior can be described with Eq. (1). Due to the non-zero $E_{gs}$, the Zeeman splitting term $g\mu_B B$ leads to a linear regime only for $B > 3 \; mT$, when the applied magnetic field is large enough to separate the two otherwise partially overlapping $v_{1,2}$ transitions.

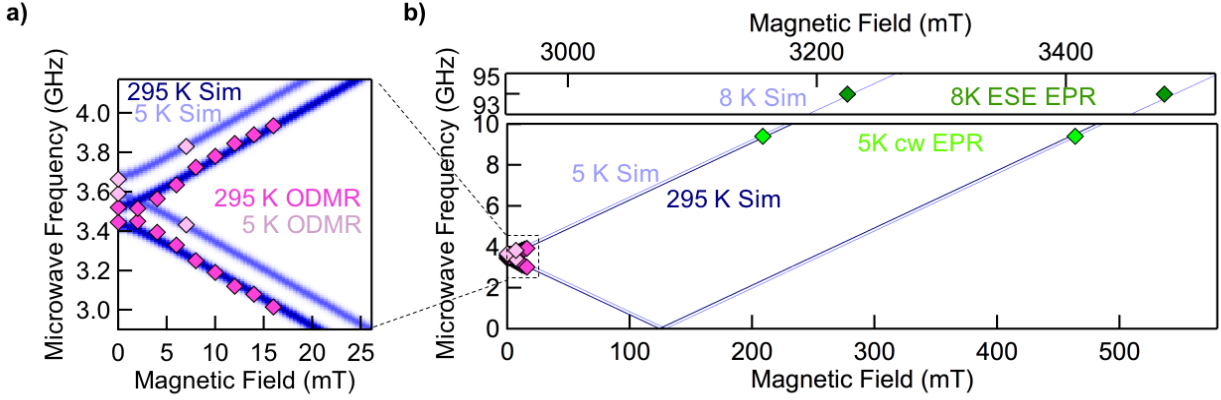

**Figure 4.** Experimental (diamonds) and simulated (dark and light blue traces) resonant frequencies of $V_B^-$ for different temperatures and magnetic fields. **a)** ODMR measurements (pink diamonds) at B < 20 mT. **b)** cw EPR measurement at T = 5 K and microwave frequency of 9.4 GHz (light green) and electron spin-echo measurements at T = 8 K and 94 GHz (dark green). Note the axes are shifted for better visibility and comparability.

To extend the magnetic field range of our measurements beyond the confocal ODMR setup limit of $\approx 20 \text{ mT}$, we applied cw electron paramagnetic resonance (cw EPR) and electron spin-echo detected (ESE) EPR. The advantage is that the cw EPR measurements are performed at a microwave frequency of $9.4 \text{ GHz}$ (X-band) (light green diamonds) and the ESE EPR measurements at a microwave frequency of $94 \text{ GHz}$ (W-band) (dark green diamonds) which allow extending the magnetic field range to $3500 \text{ mT}$. The multifrequency spin resonance approach enables us to determine the g-factor with extremely high accuracy as $g = 2.0046 \pm 0.0021$.

**Discussion**

As we mentioned, one of the crucial parameters for high-sensitivity sensing is the distance between the sensor and the object to be sensed. In this regard, sensors based on the hBN $V_B^-$ center are particularly appealing as — based on the presented model of lattice constant variations — the required hardware ideally consists of only three hBN layers, with the intermediate one hosting a $V_B^-$ center. This corresponds to a minimum distance of the defect to the surface in the sub-nanometer range of $\approx 0.33$ nm. A further thickness reduction, e.g. to a single monolayer, would indeed eliminate the interlayer contribution $c(T)$, but make the sensing effect completely dependent on the interaction of the $V_B^-$ wave function with the parameters of the adjacent material. For single layers, $D_{gs,295K}$ would differ from the theoretical calculation[29], and a calibration would be required to determine the set of parameters ($D_{gs,295 K}$, $\theta_a$, $\theta_c$) specific for the adjacent material. To substantiate this hypothesis, however, further measurements and calculations must be carried out.

In the following, we benchmark the properties of the hereby proposed sensor based on $V_B^-$ centers in hBN against other defect-based sensors in silicon carbide and diamond. For this purpose, the general sensitivity is derived, which includes the respective coupling coefficient $\Gamma_{T,\sigma,B}$ representing the sensitivity of the ODMR frequency shift due to the corresponding external influence, and the general resolution in relative change of frequency in relation to acquisition time and noise level[5]. An overview of all calculated coupling coefficients and resolutions is summarized in Table 2.

In order to facilitate comparison with other color centers in 3D materials, we consider a linear regime (50 – 350 K) in our analysis. In this range, a proportionality between $\Delta D_{gs}$ and $T$ is given by the factor $\Gamma_T = -623 \frac{kHz}{K}$. This value is almost one order of magnitude larger than the corresponding factor for NV⁻ centers in diamond ($-74 \frac{kHz}{K}$) [30]. This remarkable (> 8-fold) difference is in particular due to the larger relative change of the lattice parameters as a function of temperature in hBN, while $\theta$ is of the same order of magnitude. For NV⁻ centers, a value of $\theta_{NV} = -14.41 \text{ GHz}$ is reported[26], which is comparable to $\theta_c \approx -24 \text{ GHz}$ for $V_B^-$ in hBN. Note that $\theta_a$ can be neglected here, since on the one hand the relative change of the lattice parameter $a$ is negligible and on the other hand it counteracts the effect due to the expansion of the in-plane distance while cooling the sample. The general resolution $\delta_T^{295K}$ obtained at room temperature is approximately $3.82 \frac{K}{\sqrt{Hz}}$ which is of the same order of magnitude as defect-based temperature sensors in SiC ($1 \frac{K}{\sqrt{Hz}}$) using the same cw ODMR set-up[5]. It should be noted, however, that at cryogenic temperatures the resolution of the $V_B^-$ center is enhanced by a factor of $\approx 20$ ($\delta_T^{50K} = 0.19 \frac{K}{\sqrt{Hz}}$), as both the ODMR contrast ΔPL/PL and the PL intensity increase, which significantly reduces the required measurement time. Despite the smaller coupling coefficient, a temperature sensor based on NV's in diamond is still more sensitive ($0.76 \frac{mK}{\sqrt{Hz}}$) [31], mainly due to stronger PL emission, higher ODMR contrast and an optimized pulsed measurement protocol that exceeds the sensitivity of standard cw ODMR measurements performed here. Analogously, the magnetic field resolution of the $V_B^-$ can be quantified at room temperature as $\delta_B^{295K} = 85.1 \frac{\mu T}{\sqrt{Hz}}$ ($\delta_B^{50K} = 4.33 \frac{\mu T}{\sqrt{Hz}}$ at T=50 K). This is comparable to $V_{Si}$ in SiC ($10 \frac{\mu T}{\sqrt{Hz}}$) [5] but lower than for NVs in diamond ($3 \frac{nT}{\sqrt{Hz}}$) [32]. However, spin defects in 3D materials can lose their superior properties if they are close to the surface of the crystalline host[10], which also leads to an inevitable limitation of the minimum achievable sensor-to-object distance. This can significantly hinder the effectiveness of the spin defects-based sensors in 3D materials, highlighting the potential advantages of the $V_B^-$ center as a sub-nanometer scale sensor.

|  | Coupling coefficient Γ | | | Resolution δ (295 K) | | | Resolution δ (50 K) |
|---|---|---|---|---|---|---|---|
|  | hBN | diamond | SiC | hBN | diamond | SiC | hBN |
| Magnetic Field B | $g\mu_B = 28.0 \frac{kHz}{\mu T}$ (for $g = 2.00$) | | | $85.1 \frac{\mu T}{\sqrt{Hz}}$ | $3 \frac{nT}{\sqrt{Hz}}$ | $10 \frac{\mu T}{\sqrt{Hz}}$ | $4.33 \frac{\mu T}{\sqrt{Hz}}$ |
| Temperature T | $-623 \frac{kHz}{K}$ | $-74 \frac{kHz}{K}$ | $-1.1 \frac{MHz}{K}$ | $3.82 \frac{K}{\sqrt{Hz}}$ | $0.76 \frac{mK}{\sqrt{Hz}}$ | $1 \frac{K}{\sqrt{Hz}}$ | $0.19 \frac{K}{\sqrt{Hz}}$ |
| X Pressure $\sigma_x$ | $-0.136 \frac{Hz}{Pa}$ | | | $17.5 \cdot 10^6 \frac{Pa}{\sqrt{Hz}}$ | | | $0.891 \cdot 10^6 \frac{Pa}{\sqrt{Hz}}$ |
| Y Pressure $\sigma_y$ | $-0.212 \frac{Hz}{Pa}$ | | | $11.2 \cdot 10^6 \frac{Pa}{\sqrt{Hz}}$ | | | $0.572 \cdot 10^6 \frac{Pa}{\sqrt{Hz}}$ |
| Z Pressure $\sigma_z$ | $-0.91 \frac{Hz}{Pa}$ | | | $2.62 \cdot 10^6 \frac{Pa}{\sqrt{Hz}}$ | | | $0.133 \cdot 10^6 \frac{Pa}{\sqrt{Hz}}$ |

*Table 2. Three spin hosting systems in comparison: coupling coefficient Γ and general resolution δ at room (T = 295 K) and cryogenic (T = 50 K) temperatures. Also shown are the reference values for spin defects in diamond[30-32] and SiC[5], respectively (see main text).*

**Conclusion**

In this work, we have analyzed the spin properties of $V_B^-$ lattice defects in van der Waals hBN crystals in terms of their sensitivity to external perturbations and evaluated their advantages and disadvantages for possible applications as a nanoscale quantum sensor. The advantages include the simple intrinsic nature of the defect basically consisting of a missing boron atom, but also the potentially accessible very small distance between the sensor and the object to be sensed. In particular, we focused on the influence of temperature on the ground-state zero-field splitting, which can be directly measured by cw ODMR and is explained by the temperature-dependent lattice compression/expansion. Externally applied pressure can also induce lattice deformations and therefore be mapped onto the defect ODMR spectrum of the $V_B^-$. However, temperature and pressure measurements exclude each other and need to be performed isobar or isothermal, respectively. Nevertheless, $V_B^-$ can be used for simultaneous magnetic field measurements with high sensitivity, due to the invariability of its g-factor with respect to temperature and pressure. By comparing three spin defect hosting solid systems, diamond, SiC and hBN, we showed that the $V_B^-$ defect has comparable and, in some cases, even superior properties compared to 3D hosts. The coupling coefficient between zero-field splitting and temperature is eight times larger than the corresponding factor for NV centers in diamond. For completeness, a temperature sensor based on NV centers in diamond is still more sensitive, mainly due to stronger PL emission, higher ODMR contrast and an optimized pulsed measurement protocol that exceeds the sensitivity of cw ODMR measurements. The resolution of $V_B^-$ to external magnetic fields is comparable to that of silicon vacancies in SiC, but lower than that of NV centers in diamond. However, we believe that the recent demonstration of coherent control of $V_B^-$ spins in hBN together with the overcoming of inhomogeneous ODMR line broadening by multifrequency spectroscopy[20] will stimulate the development of advanced pulse protocols[33] and lead to a further increase in the resolution of this sensor. In addition, during preparation of our manuscript, a similar work on $V_B^-$ in hBN was submitted[34] reporting an ODMR contrast of almost 10% and its high temperature stability up to 600 K. Finally, the unique feature of hBN is its non-disturbing chemical and crystallographic compatibility with many different 2D materials, which gains a new fundamental functionality with the embedded spin centers and allows sensing in heterostructures with high sensitivity serving as a boundary itself.

## Methods:

**ODMR:** The low-field ODMR measurements are performed with a lab-built confocal microscope setup. A 532-nm laser (Cobolt Samba 100) is coupled into a 50-µm fiber and focused on the sample with a 10× objective (Olympus LMPLN10XIR), which excites an area on the sample with a diameter of about 10 µm. The photoluminescence is separated from the laser by a dichroic mirror and the remaining laser light is rejected by a 532-nm long pass filter. The photoluminescence is then coupled into a 600 µm fiber and directed onto an avalanche photodiode (Thorlabs APD440A). A 0.5-mm wide copper strip-line is used to apply microwaves to the hBN sample placed on top. Microwaves from a signal generator (Stanford Research Systems SG384) are amplified by a Mini Circuits ZVE-3W-83+ amplifier. Lock-in detection is used (Signal Recovery 7230) by on-off modulation of the microwaves. For an external magnetic field, a permanent magnet is placed below the sample.

## Acknowledgements
A.G, A.S. and V.D. acknowledge financial support from the DFG through the Würzburg-Dresden Cluster of Excellence on Complexity and Topology in Quantum Matter—ct.qmat (EXC 2147, project-id 39085490) and C.K through the DY18/13-1. I. A. and V.S. are grateful for the Alexander von Humboldt (AvH) Foundation for their generous support. The Australian Research council (via DP180100077, DP190101058 and CE200100010), the Asian Office of Aerospace Research and Development grant: FA9550-19-S-0003 are gratefully acknowledged.